\documentclass[a4paper,11pt]{article}
\usepackage[utf8]{inputenc}
\usepackage{cancel}
\usepackage{ulem}
\usepackage{amsfonts}
\usepackage{amssymb}
\usepackage{graphicx}
\usepackage{amsmath}
\usepackage{enumerate}
\usepackage{mathtools}
\usepackage{subfig}
\usepackage{color}
\usepackage{tikz}\usetikzlibrary{calc}
\usepackage{float}
\usepackage{here}
\usepackage{cite}
\usepackage{mathrsfs}
\usepackage{float,epsfig}
\usepackage{dcolumn}
\usepackage{multirow}
\usepackage{graphicx}
\usepackage{bm}
\usepackage{amsmath,amssymb,amsthm}
\usepackage[colorlinks=true,linkcolor=blue,citecolor=red]{hyperref}
\newcommand{\nn}{\nonumber}

\makeatother

\newcommand{\be}{\begin{equation}}
\newcommand{\ee}{\end{equation}}
\newcommand{\bea}{\begin{eqnarray}}
\newcommand{\eea}{\end{eqnarray}}
\numberwithin{equation}{section}
\newcommand{\Lim}{\displaystyle\lim}
\begin{document}
\title{\normalsize
\phantom{fff}

\vspace{1cm}
{\bf \Large Rotating reduced  Kiselev black holes:  \\Shadows,  Energy emission and Deflection of light  }}
\author{ M. Benali$^{1}$\thanks{mohamed.benali@um5r.ac.ma},  A. El Balali   $^{1}$ \thanks{anas.elbalali@gmail.com}, 
\hspace*{-4pt} \\
{\small $^1$
Département de Physique, Equipe des Sciences de la matière et du rayonnement, ESMAR
}\\{\small Faculté des Sciences, Université Mohammed V de Rabat, Rabat, Morocco }\\
} \maketitle

	\begin{abstract}
		{\noindent}
In this paper, we generate a rotating solution of the reduced Kiselev black hole through the modified Newman-Janis formalism. Based on such solution, we remark  different shadow behaviors by varying   the involved parameters $r_k, a, \alpha$. Concretely, we observe that the allowed values of the spin parameter $a$ are much less than the usual rotating black holes. By deeply analysing the shadow shapes, we show that comparable shadow shapes emerge for the same ratio $a/r_k$. On the other hand, we recognize that the parameters $a$ and  $\alpha$ governs the shadow geometry while the parameter $r_k$ rules the size of such a quantity. Besides, we notice that an elliptic shadow geometry appears   for  certain range of relevant parameters. By making contact with the observational side, we provide a constraint  on the  rotating reduced Kiselev (RRK) black hole parameters. In particular, we find a good compatibility between the theoretical  and experimental results.  Regarding Hawking radiation, we note that the Kiselev radius $r_ k$ shows a similar behavior to the quintessence filed intensity $\mathbf{c}$. Concerning the light motion in the vicinity of a RRK black hole, we investigate deeply the deflection by varying the relevant  parameters. In particular, we remark that such a quantity decreases by increasing the parameters $a$ and $\alpha$ while the opposite effect is observed when increasing $r_k$.

	\end{abstract}

\textbf{Keywords:} Black holes, Shadow, Deflection angle, Dark Energy.
\newpage

\tableofcontents

\newpage

\section{Introduction}

Due to its intriguing and challenging aspects, physicists have been particularly interested in black holes  for a long time. The carried researches have been considered as a crucial chance to widen our knowledge of the basic principles of physics with a special interest towards general relativity and quantum mechanics. In the study of black holes, vital explanations about the nature of the univers emerge. Thus, many inspections have been conducted to explain the enigmatic phenomena in the heart of black hole physics. In addition, the existence of these mysterious objects was recently proven through images captured by the Event Horizon Telescope (EHT) \cite{I1,I2,I3}.  Moreover, it have been considered as a significant achievement of Einstein's gravitation theory \cite{E}. These images gave an unprecedented opportunity for scientists around the world to explore the details of these cosmic objects\cite{I4,I5,I6,I7,I8,I9,I10,I11,I12,I13,I14,II1,II2,II3,II4}.

Many areas of black hole physics can be examined. Essentially, the shadow of the black hole can constrain the alternative gravitational theories and is especially rich in information on the mechanics of such celestial objects \cite{I15}. Additionally, black holes deflect light because of their strong gravitational attraction, which offers a special perspective for research on these enormous cosmic objects. With the use of such aspect, scientists can explore the gravitational environment of this object and test Einstein's theory \cite{d7,d8,d9,d10,d11,Be1,d12,d13,ds13,ds14,add01,add02,Virbhadra1,Virbhadra2}. As a result, the optical along with the thermodynamic aspects of these objects have been the focus of numerous studies \cite{d17,d19,d20,d22,d24,d25,d26,d27,d28,d29,d30,add1,add2,add3,add4,add5,add6,add61,add62,ZAA1,ZAA2,Hindd2,Hindd1}. More specifically, by adopting a thermodynamic  interpretation of the cosmological constant, an investigation of the optical and thermodynamical   aspect  of charged, rotating and non-rotating black holes  was carried \cite{t1,t2,t3,add7}. Besides, such cosmological constant  has been used in order to analyze the  stability and phase transitions, observed in multiple types of black holes present in supergravity theories.  These studies  have been expended by considering the existence of a real  scalar field modeling the dark energy  and dark matter effects. Indeed, the latter, which makes up around 70\% of the total matter in the universe, has been a favorite candidate to interpret the universe expansion. This form of energy exerts a negative pressure that promotes the cosmic expansion \cite{t4,t5,t6,add8,t7}. More precisely,  the quintessential scalar field  can be considered  as a candidate to describe the dark energy. It is defined as a real, spatially homogeneous scalar field characterized by an intensity $\mathbf{c}$\cite{t8,t9}. In this context,  an equation of state  $p = \omega\rho$ constraint the
 pressure $p$ and the energy density $\rho$. In addition, the state parameter  $\omega$  being  the ratio of the pressure to the energy density describes different models depending on its value. This parameter could take a specific valus  depending on the  constraint $-1 < \omega < -1/3$\cite{t10,t11,t12}. In this way, the state parameter $\omega$ being  the ratio of the pressure to the energy density describes different models depending on its value. For instance, the state parameter $\omega=-1$ is assigned to the cosmological constant model while the case $-1/3 < \omega < 0$ can be associated to different models, i.e quintessence or K-essence \cite{DE1,DE2,DE3}.

Moreover, the quintessence field  could be considered as  a part of Kiselev solution \cite{anas-K1}. In particular, various studies have been carried out on the optical  and  thermodynamical of Kiselev black holes \cite{t13,t14,t15,t16}. Looking more closely at the Kiselev model, one could consider a positive tangential pressure value in which the  $r_g$ representing the mass term is set zero and $-1/3 < \omega < 0$ \cite{main}. In this situation, the model is so-called reduced Kiselev black hole and  the event horizon radius of  this model exists at Kiselev radius $r_k$. This type of black hole is different from the Schwarzschild black hole, where such a difference has been explained by the existence of a gravitational potential. 

In this work, the main aim is to advance in the area of black hole physics by exploring the physics behind the RRK case. Concretely, we inspect the optical characteristics of the RRK black holes by varying  the parameters $r_k, a, \alpha$. Particularly, we study the geometry by investigating the ergosphre and horizons regions. Then, we investigate the shadow while comparing our results to the Kerr and quintessential black holes. Indeed, the shadow graphs demonstrate that the RRK black hole can match precisely the Kerr solution. Eventually, the emission rate and the deflection of light are  examined for various parameters defining the RRK black hole.

The paper is organized in the following structure. Section \ref{II} is devoted to the determination of the rotating solution  by using  the  Newman-Janis formalisme.  In section \ref{III}, we explore the optical behaviors  and energy aspects associated with the  solution. Indeed, in \ref{III-1} we construct the equations of motion and explain the different shadow aspects of the RRK solution according to the involved parameters. Then, we constrain the later by comparing our results to the observation data. After exploring the Hawking radiation in section \ref{III-3}, we investigate the deviation of light by the RRK solution. In this work, we use the units in which the light speed $c$, the reduced Planck constant $\hbar$ and Newton's constant $G$ are set as $c=G=\hbar=1$.

\section{Rotating solution of the reduced Kiselev black hole}
\label{II}
In this section, we provide the rotating solution of the reduced Kiselev black hole using the Newman-Janis method without complexifcation for a general static and spherically symmetric metric. Such a procedure, was developped to avoid the complexification step known to provide a non-unique final solution \cite{n1,n2,n3}. We then present the necessary tensors and discuss the relevant geometric characteristics of the RRK solution.
\subsection{Newman-Janis procedure without complexification}
In order to determine the RRK solution, we consider the general case of the statical and spherically symmetrical space-time 
\begin{equation}
ds_{RRK}^2=-F(r)dt^2+G(r)^{-1}dr^2+H(r)d\Omega_{s}^2,
\label{Met1}
\end{equation}
where $d\Omega_{s}^2=d\theta^2+\sin^2 \theta d\phi^2$. First, with the use of the  Eddington-Finkelstein coordinates $\left(u, r, \theta, \phi \right)$, the metric could be transformed as follows
\begin{equation}
ds_{RRK}^2=-F(r) du^2-2\sqrt{\frac{F}{G}}dudr+H(r)d\Omega^2,
\end{equation}
where 
\begin{equation}
dv=dt-\frac{dr}{\sqrt{F\,G}}.
\end{equation}
In this way, the inverse metric and  the tetrad vectors are expressed in terms of the null tetrad  $Z_\alpha^\mu=(l^\mu,n^\mu,m^\mu,\bar{m}^\mu)$ as
\begin{eqnarray}
g^{\mu\nu} & = & -l^\mu n^\nu -l^\nu n^\mu +m^\mu \bar{m}^\nu +m^\nu \bar{m}^\mu, \\
l_\mu l^\mu & = &   n_\mu n^\mu = m_\mu m^\mu = l_\mu m^\mu = n_\mu m^\mu =0.
\end{eqnarray}
Thus, we find the following tetrad vectors  
\begin{equation}
l^\mu=\delta^\mu_r, \hspace{0.2cm} n^\mu=\sqrt{\frac{G}{F}}\delta^\mu_u-\frac{G}{2}\delta^\mu_r, \quad m^\mu=\frac{1}{\sqrt{2H}}\left(\delta^\mu_\theta+\frac{i}{\sin\theta}\delta^\mu_\phi\right).
\end{equation}
Concretely, the associated vectors satisfy the following  relation 
\begin{equation}
\label{ }
l_\mu n^\mu  = - m_\mu \bar{m}^\mu =-1.
\end{equation}
Using the complex notation,  we apply a mathematical transformation on the $r-u$ plane of the static metric, which results in a rotating black hole metric 
\begin{equation}
r\rightarrow r'=r+ia\cos\theta, \hspace{0.3cm} u\rightarrow u'=u-ia\cos\theta.
\end{equation}
The new tetrad vectors become
\begin{eqnarray}
l'^\mu&=&\delta^\mu_r,\\
n'^\mu&=&\sqrt{\frac{B(r,\theta)}{A(r,\theta)}}\delta^\mu_u-\frac{B(r,\theta)}{2}\delta^\mu_r,\\
m'^\mu&=&\frac{1}{\sqrt{2C(r,\theta)}}\left(ia\sin\theta(\delta^\mu_u-\delta^\mu_r)+\delta^\mu_\theta+\frac{i}{\sin\theta}\delta^\mu_\phi\right),
\end{eqnarray}
where  the functions $\left\lbrace F(r),G(r),H(r)  \right\rbrace$ are replaced by $\left\lbrace A(r,\theta),B(r,\theta),C(r,\theta)  \right\rbrace$. A close examination reveals that the new metric in advanced null coordinates can be obtained using its inverse version. In fact, the revised metric is now given by
\begin{eqnarray}
ds^2&=&-Adu^2-2\sqrt{\frac{A}{B}}dudr+2a\sin^2\theta\left(A-\sqrt{\frac{A}{B}}\right)du d\phi+2a\sqrt{\frac{A}{B}}\sin^2\theta drd\phi \nonumber \\
& &+C d\theta^2+\sin^2\theta\left[C+a^2\sin^2\theta\left(2\sqrt{\frac{A}{B}}-A\right)\right]d\phi^2,
\label{eq:null_coordinate_metric_1}
\end{eqnarray}
where 
\begin{equation}
du=dt'+\gamma(r)dr, \hspace{0.5cm} d\phi=d\phi'+\beta(r) dr.
\label{eq:transformation_to_BL}
\end{equation}
An analysis reveals that the complexification sequence of the Newman-Janis method  could be solved by introducing new  real functions\cite{NJ,n1,n2,n3}. These functions satisfy to   the following constraint 
\begin{equation}
\label{ }
\lim_{a\to 0}D(r,\theta,a)=B(r), \hspace{0.3cm}  \lim_{a\to 0}E(r,\theta,a)=A(r), \hspace{0.3cm}  \lim_{a\to 0}\Psi(r,\theta,a)=C(r),
\end{equation}
where 
\begin{equation}\label{n7}
\gamma(r)=- \frac{(H\sqrt{F}+a^2\sqrt{G})}{\sqrt{G}(FH+a^2)},\,\beta(r)=-\frac{a}{FH+a^2},
\end{equation}
and 
\begin{equation}
\label{9}
A(r,\theta)=\frac{\sqrt{G}(FH+a^2\cos^2\theta) \Psi }{(H\sqrt{F}+a^2\sqrt{G} \cos^2\theta)^2},\, B(r,\theta)=\frac{FH+a^2\cos^2\theta}{\Psi}.
\end{equation}
In this setup, the function $\Psi(r,\theta,a)$  is still unknown. However, the explicit expression of this function could be determinate by using the following  constraint 
\begin{align}
\label{b1}&(I+a^2 x^2)^2 (3\frac{\partial \Psi}{\partial r}\frac{\partial \Psi}{\partial x^2} -2\Psi \frac{\partial \Psi}{\partial (r x^2)}) =3a^2\frac{\partial I}{\partial r}\,\Psi^2,\\
&[(\frac{\partial I}{\partial r})^2+I(2-\frac{\partial^2 I}{\partial r^2})-a^2x^2(2+\frac{\partial^2 I}{\partial r^2})]\Psi\nn\\
\label{b2}&+(I+a^2x^2)(4x^2\frac{\partial \Psi}{\partial x^2}-\frac{\partial I}{\partial r}\frac{\partial \Psi}{\partial r})=0,
\end{align}
where
\begin{eqnarray}
I(r) & \equiv & H(r)\sqrt{\frac{F(r)}{G(r)}}, \\
x & = & \cos\theta.
\end{eqnarray}
Concretely,  Eq.(\ref{b1}) is obtained by imposing $G_{r,\theta}=0$ with $G_{r,\theta}$ being the tensor  component of the Einstein equations. However,  Eq.(\ref{b2}) is derived by the help of the filed equation $G_{\mu\nu}=T_{\mu\nu}$ (see the appendix of ref\cite{n1}). After the calculations, the associated function  $\Psi$ is written as
\begin{equation}
\label{Psi}
\Psi=r^2+a^2\cos^2\theta.
\end{equation}
Taking the expression of the involved functions $A$, $B$ and $\Psi$,   the line element of the metric in the Boyer-Lindquist coordinates associated with RRK black hole  is  

\begin{equation}
ds^2 =- (1-\frac{f}{\Psi})dt^2+\frac{\Psi}{\Delta_{RRK}}dr^2+ \Psi d\theta^2+\frac{\Sigma\sin^2\theta}{\Psi}d\phi^2-\frac{2af\sin^2\theta}{\Psi}dtd\phi.
\end{equation}
The  reduced function  terms associated with the solution  are expressed as follows
\begin{equation}
\label{ }
f=\frac{r_k^\alpha}{r^{\alpha-2}}, \hspace{0.3cm} \Delta_{RRK}=r^2+a^2-f,  \hspace{0.3cm} \Sigma=(r^2+a^2)^2-a^2\Delta_{RRK}\sin^2\theta,
\end{equation}
At this point, $\alpha$ is linked to the state parameter  $\omega$ by the following equation
\begin{equation}
\label{ }
\alpha=3\omega+1,
\end{equation}
where $r_k$ being the  Kiselev radius.  In particular,  for $\alpha=1$ and $r_k=2M$ we get the line element of Kerr black hole solution.  In our investigation, we consider $0<\alpha<1$ which is associated to the range $-1/3<\omega<0$. In this case, the quintessence model behaves as quintessence energy while the case $-1<\omega<-1/3$ behaves as phantom energy. Since the new generated solution is completely different from the reduced-Kiselev black hole of the metric \eqref{Met1}, the Einstein tensor still needs to be determined. It is noted that, the modified Newman-Janis procedure is a method used to transform a non-rotating solution to a rotating one. However, the obtained solution does not have the same tensors as the non-rotating case. As a result, we report in this part the Einstein tensor associated with the RRK black hole. Indeed, after computations and simplifications, it can be shown that the non vanishing Einstein tensor components are
\begin{align}
G_{rr} &= \frac{2 (\alpha -1) r^2 r_k^{\alpha }}{\left(a^2 \cos (2 \theta )+a^2+2 r^2\right) \left(\left(a^2+r^2\right) r^{\alpha }-r^2 r_k^{\alpha }\right)},  \\
G_{\theta \theta} &=-\frac{(\alpha -1) r^{-\alpha } r_k^{\alpha } \left(a^2 (\alpha -2) \cos ^2(\theta )+\alpha  r^2\right)}{a^2 \cos (2 \theta )+a^2+2 r^2}, \\
G_{\phi t} &=\frac{2 a (\alpha -1) \sin ^2(\theta ) r^{-2 \alpha } r_k^{\alpha } \left(\left(a^2+r^2\right) r^{\alpha } \left((\alpha -2) a^2 \cos (2 \theta )+(\alpha -2) a^2+2 (\alpha +2) r^2\right)-4 r^4 r_k^{\alpha }\right)}{\left(a^2 \cos (2 \theta )+a^2+2 r^2\right)^3}, \\
&=G_{t\phi}, \nonumber \\
G_{tt} &=\frac{(\alpha -1) r_k^{\alpha }}{2  r^{2 \alpha } \left(a^2 \cos (2 \theta )+a^2+2 r^2\right)^3} \\
& \times  \left[ r^{\alpha } \left((\alpha -2) a^4 \cos (4 \theta )-(\alpha -2) a^4+4 \alpha  a^2 r^2 \cos (2 \theta )-4 (\alpha +4) a^2 r^2-16 r^4\right)+16 r^4 r_k^{\alpha }\right], \nonumber \\
G_{\phi \phi} &=-\frac{2 (\alpha -1) \sin ^2(\theta ) r_k^{\alpha }}{ r^{2 \alpha }\left(a^2 \cos (2 \theta )+a^2+2 r^2\right)^3} \\
&  \times \left[ \left(a^2+r^2\right) r^{\alpha } \left((\alpha -2) a^4+a^2 \cos (2 \theta ) \left((\alpha -2) a^2+(\alpha -4) r^2\right)+3 \alpha  a^2 r^2+2 \alpha  r^4\right) \right. \nonumber \\
&\qquad \left. -4 a^2 r^4 \sin ^2(\theta ) r_k^{\alpha }\right] \nonumber
\end{align}
It must be clarified that the modified Newman-Janis procedure pemits the introduction of symmetry feature and more physical reasons in addition to avoiding the complexification process. Moreover, it has been remarked that such a modified method is effective in cases where the original Newman-Janis process has failed.

\subsection{Horizon geometry and ergosphere}
\begin{figure*}[!ht]
		\begin{center}
{\small  \includegraphics[scale=0.35]{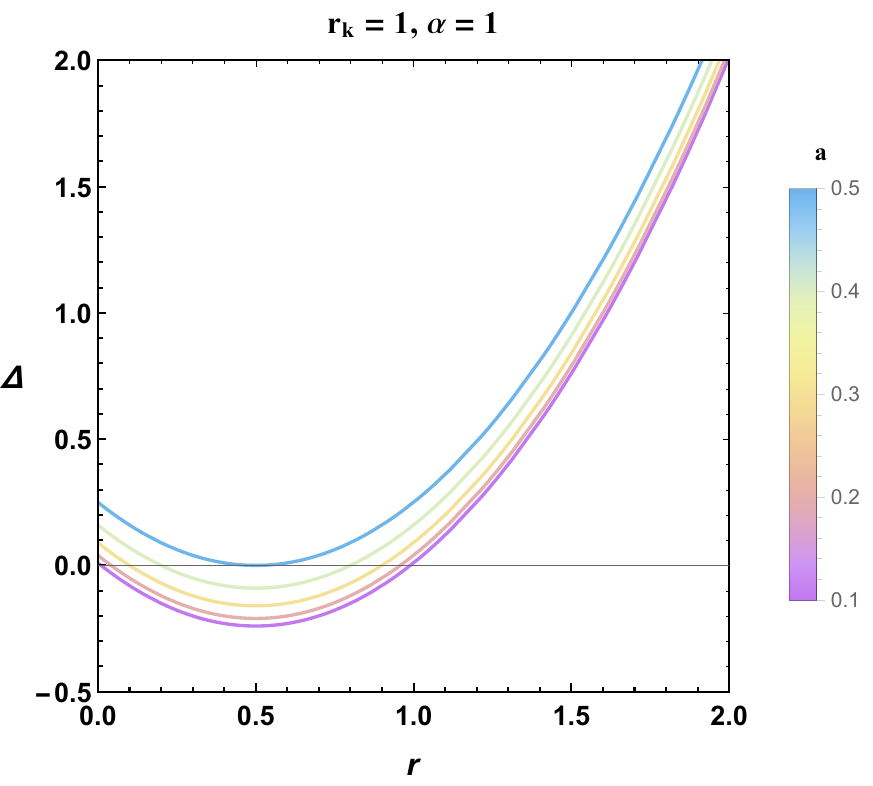}}
{\small  \includegraphics[scale=0.34]{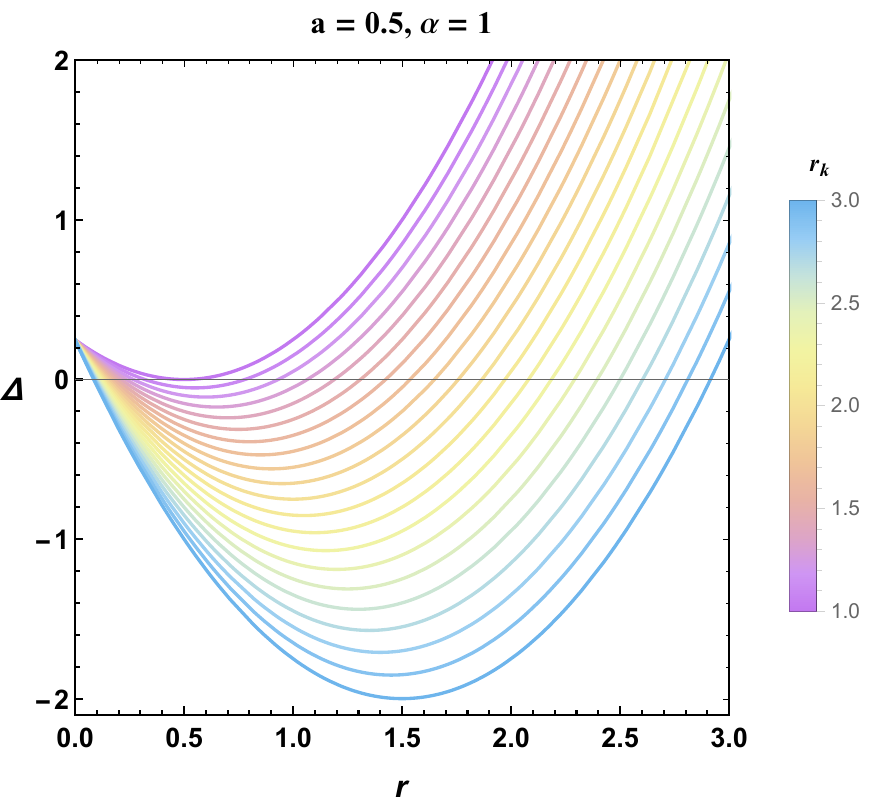}}
{\small  \includegraphics[scale=0.35]{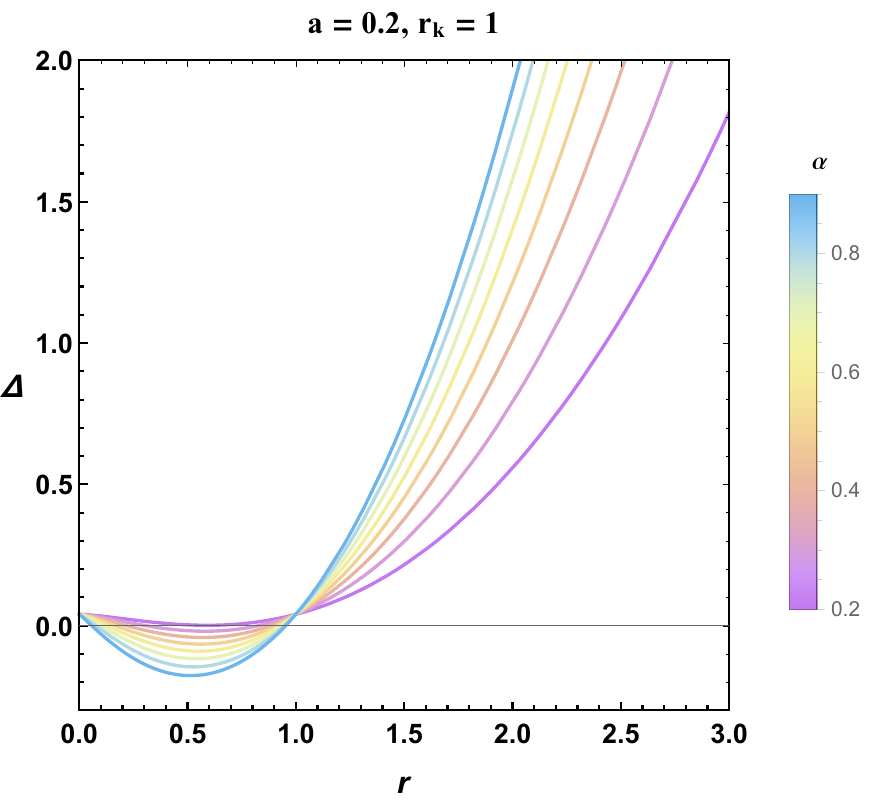}}
\caption{\it Delta metric function of RRK  black hole  for different values of the spin, Kiselev radius and $\alpha$ parameters.}
 \label{Delta}
\end{center}
\end{figure*}

To inspect the RRK black hole geometry, it is useful to recall that the solution of $\Delta=0$ provides information about the horizon radius. For example, considering $\alpha=1$, we can easily determine the inner and outer horizons that are located at
\begin{equation}
r_-=\frac{1}{2} \left(r_k-\sqrt{r_k^2-4 a^2}\right), \quad r_+=\frac{1}{2} \left(r_k+\sqrt{r_k^2-4 a^2}\right).
\end{equation}
It is clear from such equations that the Kerr black hole horizons are obtained by replacing $r_k=2M$. In the extremal case, the inner and outer horizon coincide $r_-=r_+$ and such a case is equivalent to setting $r_k=2a$. To study the aspects  of the RRK black hole  geometry, we illustrate in figure \eqref{Delta} the metric function $\Delta$ with respect to the radial coordinate  $r$ for various values of the concerned parameters.
\\
From such a figure, it can be confirmed that the RRK black hole is characterized by an inner and outer horizon for most cases. The remaining ones are associated with the extremal case in which the inner and outer horizon coincide. For the left panel, we vary the rotation parameter $a$. Clearly, we observe for $r_k=1$ and $\alpha=1$ that the extremal case is reached for $a=0.5$ while for lowest values of such parameter the black hole exhibits two horizons. For the middle panel, it can be noticed that  the gap between the inner and outer horizon increases  with $r_k$. For the $\alpha$ effect, we see that the inner horizon gets smaller when such a parameter is increased while the outer horizon shows the opposite effect. \\
In order to fully explore the geometry of the RRK solution, we examine  now the ergosphere region which is  confined by the event horizon and a static limit surface. Such region is localized outside the black hole. In fact, the ergoregion correspond to the region where the Killing vector $\chi^a$ could be viewed as a space-like vector. An interesting phenomenon can occur in this particular region of space-time. In fact, a particle can remain stationary in the ergosphere and can exit the such region. For the present solution, the ergosphere region could be calculated by the help of  this following equation 
\begin{equation}
\label{ }
r^2+a^2\cos^2\theta-f=0
\end{equation}
Using the polar coordinate,  we illustrate in figure (\ref{aa1}) the ergospheres region and the horizons for different  values of the involved parameters . In particular, for $\theta=\pi$, the equation of  ergospheres and the horizons are identical.
From such a figure, we observe that the size of the plotted quantities increases with $r_k$. Besides, it can confirmed that the gap between the inner and outer horizon increases with $r_k$. In addition, we can see that the outer ergosphere becomes more stretched in by increasing $r_k$ or $a$ . Comparing the graphs, we remark by decreasing $\alpha$ that the ergosphere becomes more prolate.
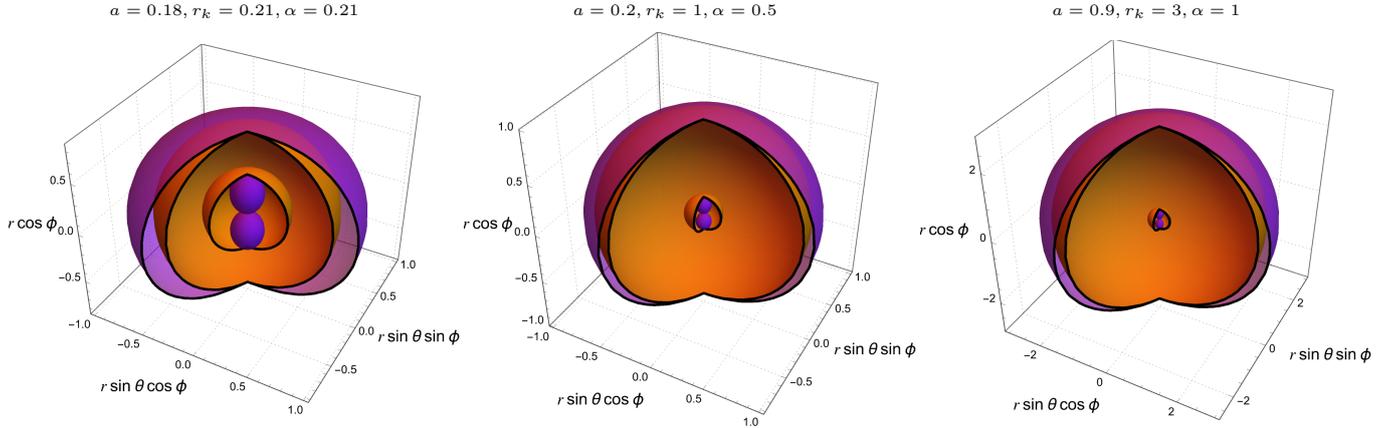
\begin{figure*}[!ht]
		\begin{center}
		\begin{tikzpicture}[scale=0.2,text centered]
		\hspace{-1.2 cm}
\node[] at (-40,1){\small  \includegraphics[scale=0.65]{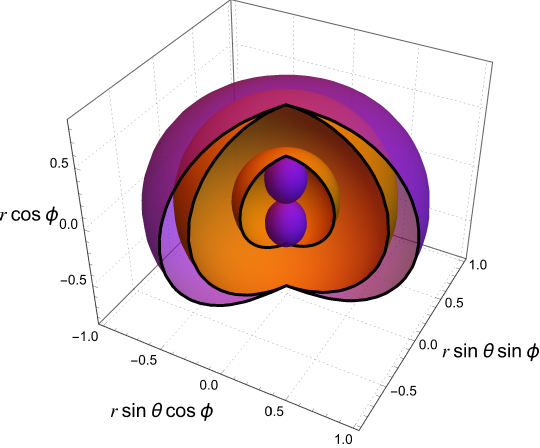}};
\node[] at (20,1){\small  \includegraphics[scale=0.65]{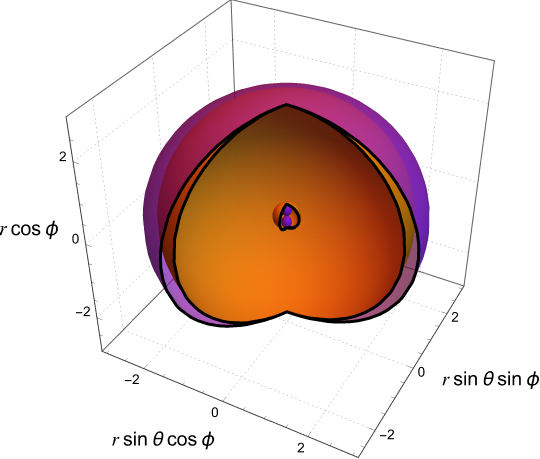}};
\node[] at (-10,1){\small  \includegraphics[scale=0.65]{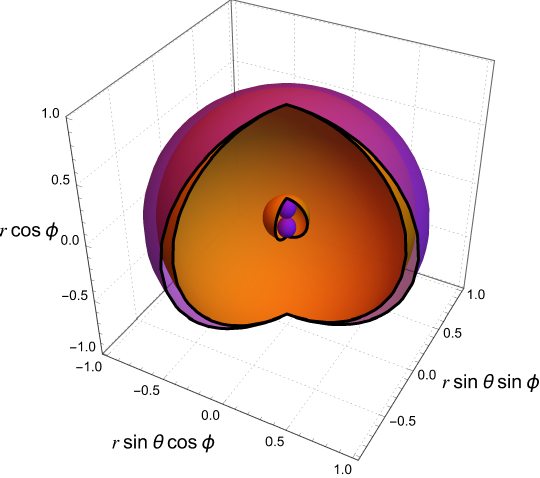}};
\node[color=black] at (-40,15.5) {\tiny$a=0.18, r_k=0.21, \alpha=0.21$};
\node[color=black] at (-11,15.5) {\tiny$a=0.2, r_k=1, \alpha=0.5$};
\node[color=black] at (20,15.5) {\tiny$a=0.9, r_k=3, \alpha=1$};
\end{tikzpicture}	
\caption{\it Ergosphere region and the horizon  variation  for different   values of the relevant parameters. The purple and orange colors are associated with the ergosphere region and the horizons, respectively.}
\label{aa1}
\end{center}
\end{figure*}
In the next setup, we analyze the optical behavior  of the present solution by varying the relevant parameters including  $\alpha$ and $r_k$.
\vspace{4cm}
\section{Shadow aspects  of RRK black hole}
\label{III}
In this section, we aim  to  study the optical aspect   of RRK solution. Concretely, we investigate the shadow behaviors and energy emission rate in terms of  the relevant parameters  associated with the present solution.  Moreover, we make contact with EHT observational data by deposing the constraint on such parameters and explore.
\subsection{Shadow aspects  of RRK black hole}
\label{III-1}
\begin{figure*}[!ht]
		\begin{center}
		\begin{tikzpicture}[scale=0.2,text centered]
		\hspace{-0.35cm}
\node[] at (-32,1){\small  \includegraphics[scale=0.56]{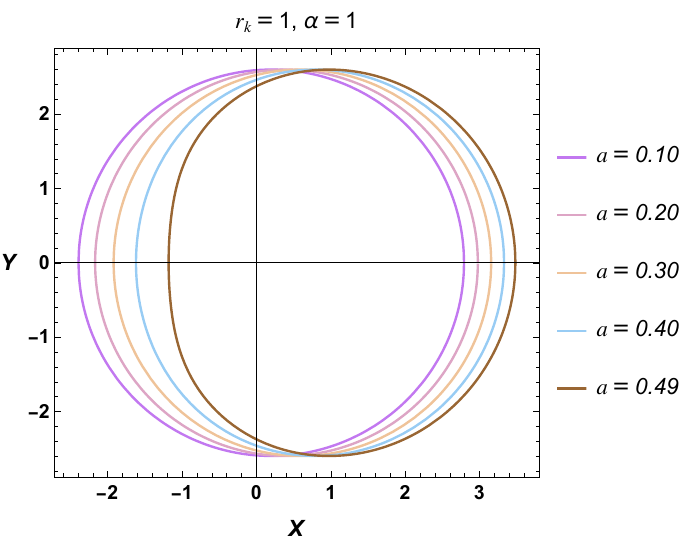}};
\node[] at (-1,1){\small  \includegraphics[scale=0.5]{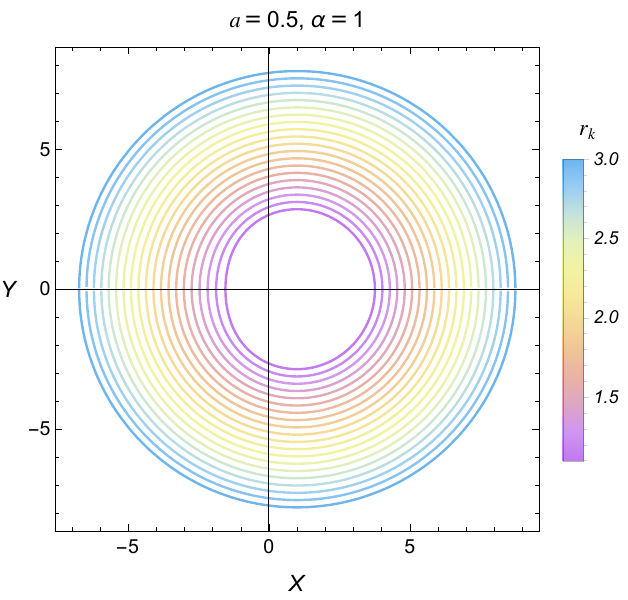}};
\node[] at (26,1){\small  \includegraphics[scale=0.5]{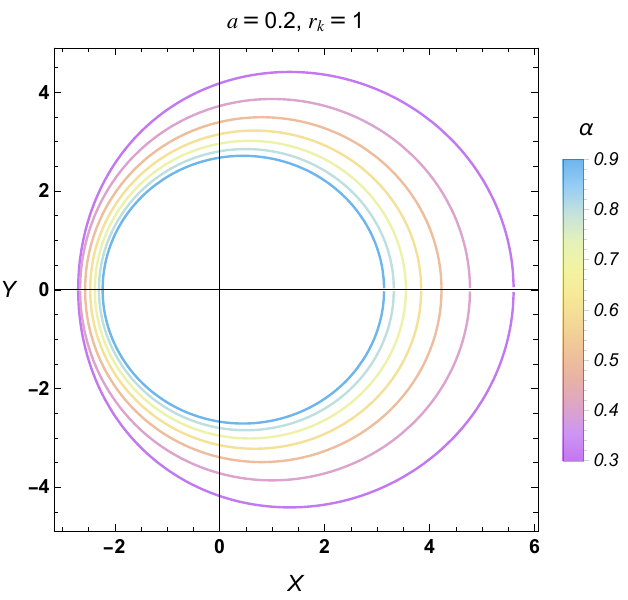}};	
\end{tikzpicture}	
\caption{Shadow behaviors  of   RRK black hole  by varying the spin, Kiselev radius and  $\alpha$ parameters.}
 \label{aas}
\end{center}
\end{figure*}

In this part, we inspect the  optical aspect of RRK black hole by  exploiting the Hamilton-Jacobi method\cite{1S1}. Indeed, this equation represents the motion of  a particle in the associated space-time which is controlled by the Jacobi action $\mathcal{S}$. In particular, for a massless particle  one  can  use 
\begin{equation}
\frac{\partial \mathcal{S}}{\partial \lambda }=-\frac{1}{2}g^{\mu \nu}p_{\mu}p_{\nu},
\end{equation}%
where $p$ and  $\lambda$  are  the four-momentum quantities and the affine parameter, respectively. Concretely, the Jacobi action in the spherically symmetric space-time has the    following form
\begin{equation}
\label{ks2}
\mathcal{S}=-Et+L\phi+S_r(r)+S_\theta(\theta),
\end{equation}
where the conserved angular momentum $L=p_\phi$ and   the conserved  total energy $E=-p_t$ are associated with  the four-momentum components of    $p_\mu$. 
Besides,  $S_\theta(\theta)$ and  $S_r(r)$  depend respectively on the angular parameter  $\theta$ and the  radial coordinate $r$. Using the  separation method, the null geodesic equations  is derived by using  the Carter mechanism \cite{1S1}. For simplicity reason, we introduce two parameters defined in terms of  the  conserved angular momentum, the total energy  and the separable constant $\mathcal{K}$
\begin{equation}
\label{xe}
 \xi_{RRK}=\frac{L}{E}, \hspace{1.5cm}\eta_{RRK}=\frac{\mathcal{K}}{E^2}.
\end{equation}
Using theses  two quantities,  the null geodesic equations  of RRK black hole  are  derived by  solving the following equations
\begin{align}
\Psi \frac{d  \, t}{d \tau}& =  E \left[ \frac{r^2+a^2}{\Delta_{RRK}}\left(r^2+a^2-a\xi\right)+a\left(\xi-a\sin^2\theta\right)\right], \\
\Psi  \frac{d  \, r}{d \tau} &=\sqrt{\mathcal{R}(r}),\\
 \Psi\frac{d  \, \theta}{d \tau} & =\sqrt{\Theta(\theta)},\\
\Psi \frac{d  \, \phi}{d \tau} &= E \left[ \frac{a}{\Delta_{RRK}}\left(r^2+a^2-a\xi\right)+\left(\frac{\xi}{\sin^2}-a\right)\right],
\end{align}
where the radial and the polar functions        are expressed as follows
 \begin{align}
\mathcal{R}(r)&=E^2\left[ \left[ \left( r^2 +a^2 \right)  -a \xi_{RRK}  \right]^2- \Delta_{RRK}(\eta+(\xi-a)^2)   \right], \\
 \Theta(\theta)&= E^2 \left[ \eta_{RRK}  - \frac{1}{\sin^2\theta} \left( a\sin^2 \theta - \xi_{RRK}\right )^2 +(\xi-a)^2\right].
\end{align}
With the use of the radial function and the conditions bellow, we can elaborate the shadow of such model
\begin{equation}\label{xx1}
\mathcal{R}(r)\Big|_{r=r_0}=\frac{d\mathcal{R}(r)}{d r}\Big|_{r=r_0}=0,
\end{equation}
where, $r_0$    is the radius of photon sphere. Various equations are used to determine this radius for different black hole solutions\cite{d17,d19,d20,d22,d24,d25,d26,d27,d28,d29,d30,add1,add2,add3,add4,add5,add6,add61,add62,Hindd1,Hindd2}. Considering  $ \Theta(\theta)>0$, the  parameters $\xi$ and $\eta$ are expressed    as a function of the involved parameters associated with the RRK black hole solution. Indeed, we obtain
\begin{align}
& \eta_{RRK} = \frac{a^2 \left((\alpha -2) r_k^{\alpha }-2 r^{\alpha }\right)+r^2 \left((\alpha +2) r_k^{\alpha }-2 r^{\alpha }\right)}{a \left(2 r^{\alpha }+(\alpha -2) r_k^{\alpha }\right)} \bigg\vert_{r=r_0},\\
& \xi_{RRK}=\frac{8 a^2 \alpha  r^{\alpha +2} r_k^{\alpha }-4 r^{2 \alpha +4}+4 (\alpha +2) r^{\alpha +4} r_k^{\alpha }-(\alpha +2)^2 r^4 r_k^{2 \alpha }}{a^2 \left(2 r^{\alpha }+(\alpha -2) r_k^{\alpha }\right)^2} \bigg\vert_{r=r_0}.
\end{align}
Taking $\alpha=1$ and $r_k=2M$, we recover the   equations of motion and the impact parameters of Kerr solution \cite{I15}. It is noted that,   the shadow computations need  certain relevant parameters. Indeed,  we introduce the celestial coordinates that control the  statical observer in the associated space-time. Thus, the   boundary of   shadow can  be  approached by  using  the celestial coordinates $X$ and $Y$\cite{I15,d9,d10}. For our solution, the  celestial coordinates in the  equatorial plan are expressed as follows  
 \begin{eqnarray}
{X}& = & -\xi_{RRK}, \\
{Y} & = & \pm \sqrt{\eta_{RRK}},
\end{eqnarray}
Considering these two coordinates, we examine the shadows behavior by varying the relavant parameters associated    with the present solution. In Fig.(\ref{aas}), we illustrate the shadow shape of  RRK black hole by varying the relavant parameters including the rotating one.  
First, we would like to get the analogie with Kerr black hole solution. To do so,  we consider $\alpha=1$ and we    vary the spin $a$ and the  Kiselev radius  $r_k$.  A close examination shows that  the shadow deformation increase with the spin  parameter $a$. From the left panel, we observe that the  size and the shape of RRK black hole is different from the Kerr solution for the considered values of  spin parameter $a$.  Indeed, it can be remarked that the  D-shape configuration appears for smaller values of spin parameters contraryto the Kerr black hole.  In parallel, for fixed value of spin parameters, the shadow size of the RRK black hole solution increases   with respect to  the Kiselev radius $r_k$ as it can be clearly seen from the middle panel. In particular, for the specific  value $r_k=2$,  the shadow of RRK is equivalent to the Kerr black hole. Increasing or decreasing this specific value conduct the shadow of RRK black hole to be either larger or smaller than the Kerr one. It is important to note that comparable shadow shapes can be found with the use of the ratio $a/r_k$. Precisely, the D-shape appears in the RRK black hole for $r_k=1, a=0.49$ while in the Kerr black hole for $r_k=2, a=0.98$. Thus, the D-shape arises for the same ratio $a/r_k$. However, the size is different due to the value of $r_k$. In the right side of the figure, we  vary  the $\alpha$ parameter for fixed value of  the Kiselev radius and the spin parameters.  Based on this figure, the shadow of RRK solution decreases by increasing   $\alpha$. However, for small values of $\alpha$ we get an elliptic  geometry of the shadow contrary to the ordinary solutions of black holes. Thus, it can be deduced from such figures that the  Kiselev radius $r_k$ controls the size of the shadow and $\alpha$ governs the geometry. This results are  perfectly  match with the previous works  associated  the   quintessential  background existence \cite{I13,L1}. Indeed, in several  works it has been shows that the quintessential field intensity rules the shadow size. As a result, we conclude that the kiselev radius $r_k$ effect on the shadow geometry  is equivalent the quintessential field intensity $\mathbf{c}$. Moreover, comparing the present results  with  the quintessential and  cosmological black hole. A  close examination show that, the $\alpha$ parameter in the  RRK black hole  could manipulated the shadow geometry contrary to the  quintessential AdS solution.

 \subsection{Shadow constraints  via the observational data}
\label{III-2}
\begin{figure*}[!ht]
		\begin{center}
		\begin{tikzpicture}[scale=0.2,text centered]
		\hspace{-0.3 cm}
\node[] at (-30,1){\small  \includegraphics[scale=0.55]{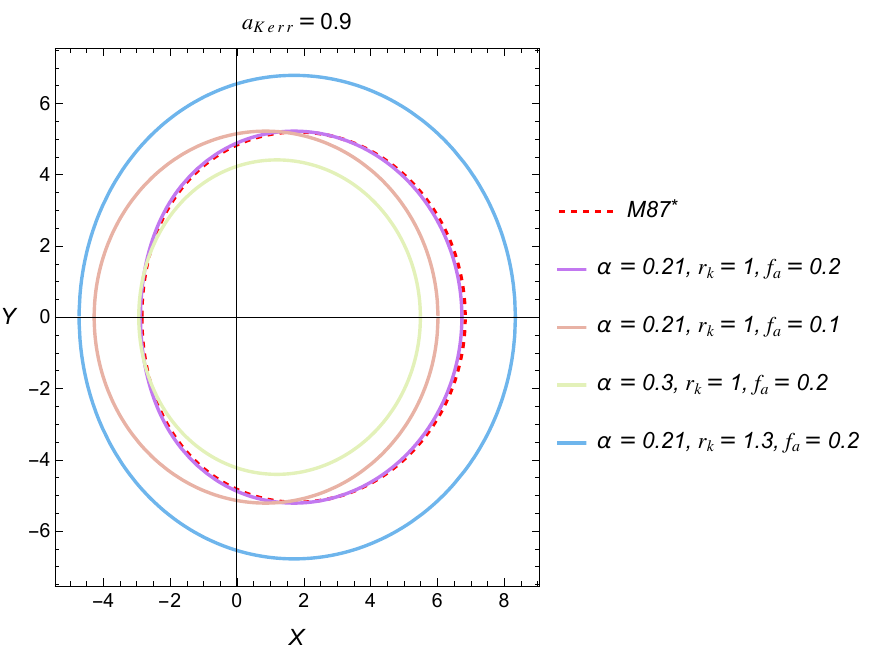}};
\node[] at (8,1){\small  \includegraphics[scale=0.7]{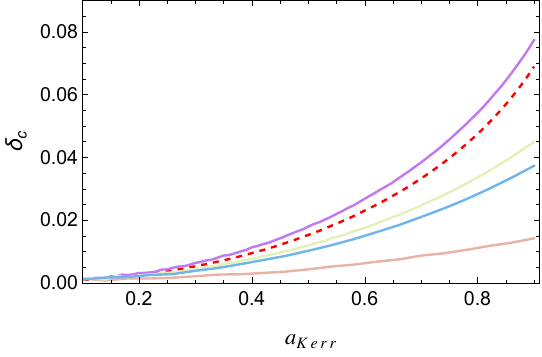}};	
\end{tikzpicture}	
\caption{\it Shadow  and the distortion of   M8$7^\ast$ and RRK black hole  for different constraint parameters.}
 \label{beal}
\end{center}
\end{figure*}

 In this part, we approach  the maximal radius of shadows $R_s$ and the geometry distortion $\delta_c$. Besides, we make contact with the EHT collaboration by imposing constraints on the involved parameters.  In the present solution, we have two different  geometry of shadow, the circular and elliptic configuration. Based on the equations reported about such works, we compute the  $R_s$ and $\delta_c$ parameters for the circular and the  elliptic configuration\cite{I7,I12,d11,Be1}. In Table.(\ref{RR1}), we calculate    $R_s$ and $\delta_c$ quantities  by varying  the relevant parameters controlling the RRK black hole solution.
\begin{table}[ht!]
\begin{center}
 \scalebox{0.9}{
\begin{tabular}{|l|ccccc|ccccc|ccccc|}
\hline
\multirow{3}{*}{}             & \multicolumn{5}{c|}{$r_k=1$ and $\alpha=1$}                                                                        & \multicolumn{5}{c|}{$a=0.5$ and $\alpha=1$}                                                                        & \multicolumn{5}{c|}{$a=0.2$ and $r_k=1$}                                                           \\ \cline{2-16} 
                              & \multicolumn{5}{c|}{$a$}                                                                                           & \multicolumn{5}{c|}{$r_k$}                                                                                         & \multicolumn{5}{c|}{$\alpha$}                                                                                        \\ \cline{2-16} 
                              & \multicolumn{1}{c|}{0.1}  & \multicolumn{1}{c|}{0.2}  & \multicolumn{1}{c|}{0.3}  & \multicolumn{1}{c|}{0.4}  & 0.49 & \multicolumn{1}{c|}{1.1}  & \multicolumn{1}{c|}{1.5}  & \multicolumn{1}{c|}{2}    & \multicolumn{1}{c|}{2.5}  & 3    & \multicolumn{1}{c|}{0.3}  & \multicolumn{1}{c|}{0.5}  & \multicolumn{1}{c|}{0.7}  & \multicolumn{1}{c|}{0.9}  & 1    \\ \hline
\multicolumn{1}{|c|}{$R_s$} & \multicolumn{1}{c|}{2.59} & \multicolumn{1}{c|}{2.59} & \multicolumn{1}{c|}{2.59} & \multicolumn{1}{c|}{2.59} & 2.59 & \multicolumn{1}{c|}{2.86} & \multicolumn{1}{c|}{3.90} & \multicolumn{1}{c|}{5.19} & \multicolumn{1}{c|}{6.50} & 7.80 & \multicolumn{1}{c|}{4.41} & \multicolumn{1}{c|}{3.45} & \multicolumn{1}{c|}{3.01} & \multicolumn{1}{c|}{2.71} & 2.60 \\ \hline
\multicolumn{1}{|c|}{$\delta_c$} & \multicolumn{1}{c|}{0.08} & \multicolumn{1}{c|}{0.15} & \multicolumn{1}{c|}{0.24} & \multicolumn{1}{c|}{0.33} & 0.44 & \multicolumn{1}{c|}{0.39} & \multicolumn{1}{c|}{0.27} & \multicolumn{1}{c|}{0.20} & \multicolumn{1}{c|}{0.15} & 0.13 & \multicolumn{1}{c|}{0.33} & \multicolumn{1}{c|}{0.24} & \multicolumn{1}{c|}{0.20} & \multicolumn{1}{c|}{0.17} & 0.15 \\ \hline
\end{tabular}}
\end{center}
\caption{Geometrical deformation  of   RRK black hole  by varying the spin, Kiselev radius and the $\alpha$ parameters.}
\label{RR1}
\end{table} 
\\
It is worth noting that  the shadow radius increase (decrease) by increasing  the Kiselev radius (increasing $\alpha$ parameter). However, $R_s$ is almost  constant  by varying the spin parameter.  Moreover, the shadow distortion increase with spin parameter and  decrease by increasing the   Kiselev radius  or $\alpha$. It is evident  that the spin parameter $a$ increase the shadow distortion like the usual rotating black holes. In the present solution, the spin and the parameter $\alpha$  controls the shape of black hole while the Kiselev radius governs its size. However, in the rotating solution with the quintessential  background, only the rotation parameter $a$ could  affect the shadow deformation\cite{I13,L1}.\\
Now, we  make contact with the EHT observational data by imposing a constraint on such parameters including the rotating one. Indeed, we rely on the observational data from international EHT collaborations. This data is  linked to  the supermassive $M87^\ast$ black hole  shadow. Moreover, the EHT collaborations data could be exploited to test and explore any  proposed models associated with the black hole solutions. Previous studies have demonstrated that 
we might impose such constraints on the relevant parameters controlling the black hole geometry\cite{I11,Be1}.
In the present solution, we constraint the  relevant parameters $a$, $r_k$ and $\alpha$.
Indeed,  by normalizing the mass of $M87^\ast$, we could compare the  shadow  of M8$7^\ast$  and RRK black holes.  Using the $M87^\ast$ black hole mass $M_{BH} = 6.5 \times10^9M_\odot$ and $r_0 = 91.2\;kpc$, we find by plotting  both  shadows that the Kiselev radius and $\alpha$ parameter should take the following values
\begin{equation}
\label{ }
r_k=1 \hspace{1cm} \text{and}  \hspace{1cm}  \alpha=0.21
\end{equation}
For the rotating parameter, we remarked that $a$  could be constrained  by using the following relation
\begin{equation}
\label{constra}
a=f_{a}\;a_{Kerr}
\end{equation}
where $f_{a}$ and $a_{Kerr}$ are  the  scalar factor and the   rotating parameter of  Kerr solution, respectively. Indeed,  we plot in Fig.(\ref{beal}) the shadows and the distortion parameter for M8$7^\ast$  and RRK black holes.
In this figure, we remarque a good compatibility between shadow geometry of RRK black hole and  $M87^\ast$ for $\alpha=0.21, r_k=1\;\text{and}\; f_a=0.2$.  As expected, the shadows size of the $M87^\ast$  and RRK solution are almost equal for the last constraint  of parameters.  Besides, we vary  the rotating parameter by using the constraint in equation (\ref{constra}). In this case, the  distortion $\delta_c$ for the experimental  and the RRK black hole  are nearly identical. Examining values that deviate from the constraints, we notice that the shadow size and distortion  are different than $M87^\ast$.
\subsection{Energy emission rate} 
\label{III-3}
In the proximity of black holes, quantum oscillations generate and annihilate a large number of pairs of particle near the horizon. This causes a tunneling effect to emit positive-energy particles outside the black hole, in the region where Hawking radiation manifests . This phenomenon, known as Hawking radiation, leads to the gradual evaporation of the black hole over a period of time. We look specifically at the energy emission rate associated with this process. For a very distant observer, the high energy absorption cross-section tends to approach the shadow of the black hole. Moreover, at very high energies, the effective absorption cross-section of the black hole oscillates until it reaches a constant limiting value $\sigma_{lim}\approx \pi R_s^2$. It is noted that this constant limiting value is roughly equivalent to the area of a photon sphere. In this way, the energy  emission rate of a certain black hole solution  can be  expressed as follows
\begin{equation}
\label{ }
\frac{d^2E(\varpi)}{d\varpi dt}=\frac{2\pi^2\sigma_{lim}}{e^{\frac{\varpi}{T_{BH}}-1}}\varpi^3,
\end{equation}
where $\varpi$ and $T$ are the photon frequency  and the black hole temperature  at outer horizon $r_+$ of the RRK solution \cite{emission1,emission2}. Indeed, this temperature could be calculated by using the following relation 
\begin{equation}
\label{ }
T_{BH}(r_+)=\Lim_{r\to r_+}\frac{1}{2\pi\sqrt{g_{rr}}}\frac{\partial \sqrt{-g_{tt}}}{\partial r}\Big{|}_{\theta=0}=\frac{r_+^{1-\alpha } {r_k}^{\alpha } \left((\alpha -2) a^2+\alpha  r_+^2\right)}{4 \pi  \left(a^2+r_+^2\right)^2}
\end{equation}
Sending  $\alpha\to 1$ and $r_k\to2$, we recover the temperature  expression of  the Kerr black hole solution.  In the Fig.(\ref{aa1}), we plot the emission rate behaviors of RRK black hole by varying the relevant parameters controlling the associated solution. 
\begin{figure*}[!ht]
		\begin{center}
		\begin{tikzpicture}[scale=0.2,text centered]
		\hspace{-0.3 cm}
\node[] at (-30,1){\small  \includegraphics[scale=0.55]{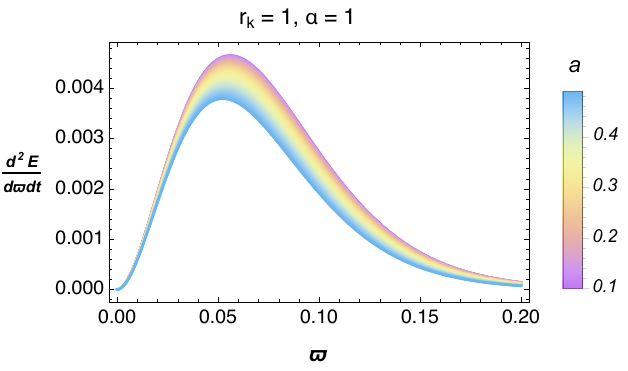}};
\node[] at (-1,1){\small  \includegraphics[scale=0.54]{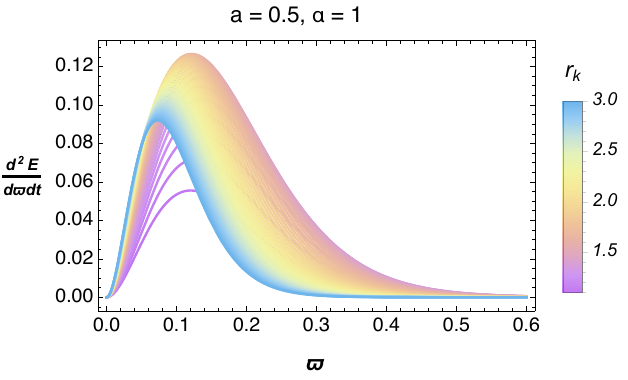}};
\node[] at (28,1){\small  \includegraphics[scale=0.54]{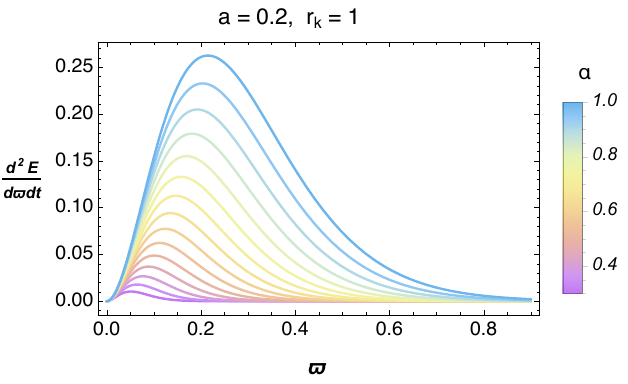}};	
\end{tikzpicture}	
\caption{\it Energy emission rate of RRK black hole by varying the relevant parameters controlled the solution.}
 \label{aa1}
\end{center}
\end{figure*}
In can be remarked that,  the  emission rate increase by decreasing the spin or  the  Kiselev radius. It is noted that the maximal values of the energy emission rate is associated with the specific values of the  frequency $\varpi_{max}$. Indeed, this value is  varying   only   with  Kiselev radius parameter. This new behavior, could be interpreted   by the fact that   the  Kiselev radius could replace   the mass parameters effect. Focusing on the variation of  the emission rate with respect to the $\alpha$ parameter, we observe that the emission rate of the RRK black hole  increases proportionally to $\alpha$. A close comparison  between the present  solution and the black hole with quintessential background reveals that,  the  Kiselev radius and the intensity of quintessential filed  $\mathbf{c}$ play the same role. Indeed, based on several works,  the intensity of quintessential filed decrease the   emission rate\cite{I13,d29,d30}.  In relation with this result, we could confirm that the Kiselev radius acts as a cooling system surrounding the RRK black hole since it is characterized by slower Hawking radiation.\\
 To complete the optical behaviors of  RRK black hole,  we examine  in the next section the deflection angle behaviors  by varying   the several  parameters including  the spin and the  Kiselev radius. 
 
 \section{Light deflection near an RRK black hole}
\label{IV}
\begin{figure*}[!ht]
		\begin{center}
		\begin{tikzpicture}[scale=0.2,text centered]
		\hspace{-0.35cm}
\node[] at (-32,1){\small  \includegraphics[scale=0.5]{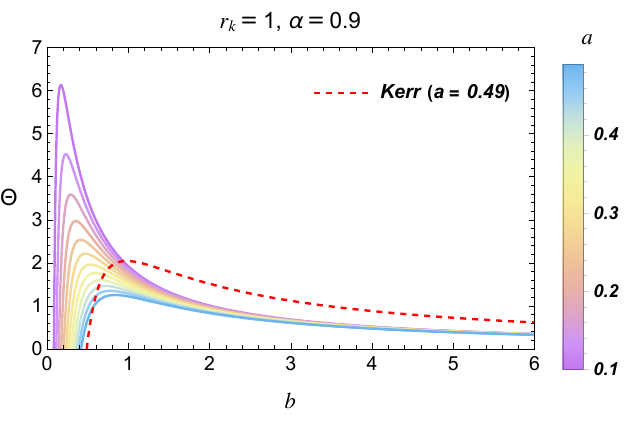}};
\node[] at (-2.5,1){\small  \includegraphics[scale=0.5]{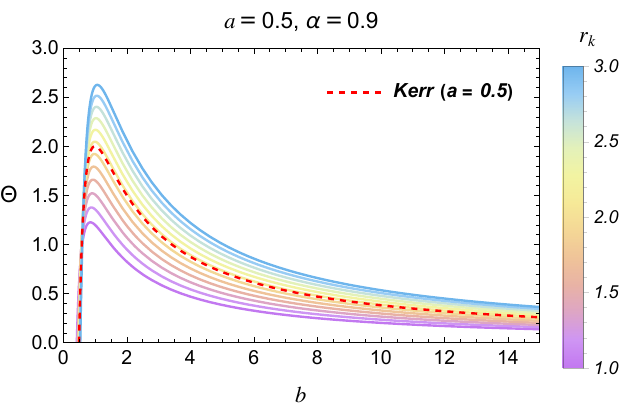}};
\node[] at (26,1){\small  \includegraphics[scale=0.5]{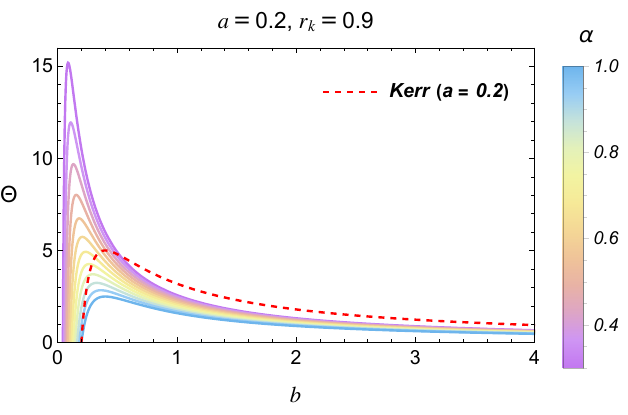}};	
\end{tikzpicture}	
\caption{\it Deflection angle behaviors of RRK  black hole  by varying the spin, Kiselev radius and the $\alpha$ parameters.}
 \label{aa}
\end{center}
\end{figure*}

In this section, we would like to examine  the light deflection  angle  by the RRK black hole solution.  An extensive examination was undertaken to explore various approaches towards    the deflection angle of the   light ray   around four-dimensional black holes. These approaches were subjected to extensive analysis with the aim of determining a considerable range of possible solutions. This particular approach is dependent on geodesic equations that determine the trajectory of massless particles. However, its application generates complex solutions characterized by elliptical function integration. Due to these complex points, one needs to  find alternative approaches to achieve more accessible or interpretable results  as part to study the  light rays motion around the compact object. In this study, a different approach was adopted in which we exploit the Gauss-Bonnet theorem based on calculations linked to optical metrics \cite{ds13,ds14}. This approach offers a promising possibility to deepen our understanding of the characteristics associated with light rays behaviors around  the  four-dimensional black holes. Concretely,  the implicit  deflection angle  expression  can be extracted using the Gauss-Bonnet formalism for ${}^{\infty}_{R}\square^{\infty}_{S}$
\begin{align}
\Theta 
=-\iint_{{}^{\infty}_{R}\square^{\infty}_{S}} K dS 
- \int_{R}^{S} \kappa_g d\ell . 
\label{}
\end{align}
where  $K$ and $ \kappa_g$ are the Gaussian  and the   geodesic curvatures, respectively. It is noted that,  the prograde case is associated with the positive values of $d\ell$. However, the opposite  is expected for   negative values of $d\ell$. To work out  the deflection angle of light ray, such cases  are needed. First,  we consider  the null geodesic condition $ds^2=0$  to get  expression of $dt$ in the following form
\begin{equation}
dt= \sqrt{\gamma_{ij} dx^i dx^j} +\beta_i dx^i , 
\label{opt} 
\end{equation}
where the metrics $\gamma_{ij}$ and  $\beta_i$ are expressed as function of the RRK black hole metric  
\begin{align}
\gamma_{ij}dx^idx^j \equiv&
\frac{\Psi ^2}{\Delta  \Psi-\Delta  f }dr^2
+\frac{ \Psi^2}{ \Psi -f}d\theta^2
+\frac{\sin ^2(\theta ) \left(\left(a^2+r^2\right)^2+\frac{a^2 \sin ^2(\theta ) \left(\Delta  \Psi -f^2-\Delta  f\right)}{f- \Psi }\right)}{ \Psi }d\phi^2 , 
\label{mohamed6}
\\
\beta_idx^i \equiv&\frac{a f \sin ^2(\theta )}{f- \Psi }d\phi . 
\label{mohamed5}
\end{align}
Indeed, we calculate the  first part of integral as a function of the relevant parameters. To do so, the Gaussian curvature $\mathbf{K}$  is needed. Concretely, in the  equatorial plan $\theta=\frac{\pi}{2}$,  $\mathbf{K}$ is expressed as function of the  Riemann tensor $R_{r\phi r\phi}$
\begin{equation}
\mathbf{K}=\frac{{}^{}R_{r\phi r\phi}}{\det({\gamma^{}_{ij}})}.
\end{equation} 
Using such order of calculation, the Gaussian curvature is expressed  as function of relevant parameters controlling the solution  
\begin{equation}
\label{mohamed2}
\mathbf{K}=-\frac{{r_k}}{r^3}+\frac{{r_k}(\alpha-1)} {2 r^3}(2 \ln (r)-2 \ln ({r_k})-3)+\mathcal{O}(r_k^3,a^2,\alpha^2,\alpha r_k^2) .
\end{equation}
In parallel,  the $dS$ is calculated by using the following  relation 
\begin{equation}
\label{mohamed3}
dS\equiv\sqrt{\det(\gamma^{}_{ij})}drd\phi=r+\frac{3\,{r_k}}{2}-\frac{3\,r_k(\alpha-1)}{2}    (\ln (r)-\ln ({r_k}))+\mathcal{O}(r_k^2,\alpha^2,\alpha^2 r_k^2).
\end{equation}
To complete the essential blocks, we need to determine the Gaussian curvature integral. In order to achive that, we employ the weak field approximations for the photon orbit equation and the slow rotation approximations. In this way, we obtain
\begin{equation}
\label{mohamed}
\left(\frac{d u}{d\phi}\right)^2=\frac{1}{b^2}-u^2+r_k\,u^3,
\end{equation} 
where $u=\frac{1}{r}$. It is  worth noting that, the photon orbit equation in the zero order could be solved by using the following solution 
 \begin{equation}
u(\phi)=\frac{\sin \phi}{b}.
\end{equation}
 In the RRK black hole,  it is necessary to obtain  this solution in the linear-order with $r_k$, $a$ and $\alpha$. To do so, we solve  the equation (\ref{mohamed}) with respect to each order, we get the  orbit equation as follow  
 \begin{equation}
\label{ }
u(\phi)=\frac{\sin \phi}{b}+\frac{r_k}{2b^2}(1+\cos^2\phi)-\frac{a\,r_k}{b^3}
+\frac{(\alpha-1)\,r_k}{4\,b^2}\,u_1(\phi)+\mathcal{O}\left(\frac{r_k^2}{4b^3},\frac{a^2}{b^3},\frac{\alpha^2\,r_k^2}{b^3}\right),
\end{equation}
where 
\begin{equation}
\begin{split}
\label{mohamed1}
u_1(\phi)=3& \ln \left(\frac{\sin \phi}{b}\right)+\cos (2 x) \ln \left(\frac{\sin \phi}{b}\right)+2 (\ln ({r_k})-1) \cos ^2\phi\\
&+2 \ln ({r_k})+4 \cos \phi \left[\ln \left(\cos \frac{\phi}{2}\right)-\ln \left(\sin\frac{\phi}{2}\right)\right].
\end{split}
\end{equation}
Up to the order $\mathcal{O}(r_k^2,a^2,\alpha^2,\alpha\,r_k^2)$, we can calculate the integral of $\mathbf{K}$ with the help of equations (\ref{mohamed2}), (\ref{mohamed3}) and (\ref{mohamed1})
\begin{align}
&-\iint_{{}^{R_{\infty}}_{R}\square^{S_{\infty}}_{S}}KdS 
=\int_{\phi_S}^{\phi_R} \int_{\infty}^{r_{OE}}\mathbf{K}\, r\,dr d\phi=-\int_{\phi_S}^{\phi_R} \int_{0}^{u(\phi)} \mathbf{K}\, u^2 du d\phi
\notag\\
=&\int_{\phi_S}^{\phi_R}
\Big[\frac{r_k}{b}\sin\phi+\frac{(\alpha-1)\,r_k}{2b} \big(\sin\phi (-2 \ln (b \csc\phi)+2 \ln ({r_k})+1)\big)\Big] d\phi
\notag\\
=&\frac{r_k}{b}\Big[\cos\phi\Big]^{\phi_S}_{\phi_R}
+\frac{r_k\,(\alpha-1)}{2b}\left(1-2\ln(r_k)\right) \Big[\cos\phi\Big]^{\phi_R}_{\phi_S}+\frac{r_k\,(\alpha-1)}{2b}\left(1-2\ln(r_k)\right) \Big[\ln(\sin\frac{\phi}{2}\cos\frac{\phi}{2})\Big]^{\phi_R}_{\phi_S}
\notag\\
=&\frac{r_k}{b}\Big[\sqrt{1-b^2{u_S}^2}+\sqrt{1-b^2{u_R}^2}\Big]-\frac{r_k\,(\alpha-1)}{2b}\left(1-2\ln(r_k)\right)\Big[\sqrt{1-b^2{u_S}^2}+\sqrt{1-b^2{u_R}^2}\Big] 
\notag\\
&-\frac{r_k\,(\alpha-1)}{b}\Big[\ln\left(\frac{\sqrt{1-b^2u_S^2}}{2}\right)-\ln\left(\frac{\sqrt{1-b^2u_S^2}}{2}\right)\Big]
\end{align}
For simplicity reasons, we have used  $\phi_R=-\sqrt{1-b^2u_S^2}+\mathcal{O}\left(\frac{r_k}{b}\right)$ and $\phi_S=\sqrt{1-b^2u_S^2}+\mathcal{O}\left(\frac{r_k}{b}\right)$. Having computed the Gaussian curvature  integral, we move to the computation of the geodesic curvature integral $\kappa_g$. Indeed, we examine the geodesic curvature of the photon's orbit in the equatorial plane. Recalling that the space associated with the generalized optical metric is axisymmetric and stationary, we get
\begin{equation}
\kappa_g=-\sqrt{\frac{1}{\det(\gamma_{ij})\gamma^{\theta\theta}}}\frac{\partial \beta_{\phi}}{\partial r}.
\end{equation}
Using the equations (\ref{mohamed6}) and (\ref{mohamed5}), the geodesic curvature could be calculated in term of $r_k$, $\alpha$ and rotating parameter
\begin{equation}
\label{ }
\kappa_g=-\frac{a\,{r_k}}{r^3}-\frac{a\,{r_k}(\alpha-1)}{r^3}\left(\ln ({r_k})-\ln (r)+1\right)+\mathcal{O}(r_k^2,\alpha^2).
\end{equation}
In order to calculate  the integral, we consider the prograde scenario in which the angular momentum associated with the photon orbits  is linearly aligned with the black hole spin. To concretize this, we adopt a linear approximation of the photon orbit as follows $r=\frac{b}{\cos v}+\mathcal{O}(r_k,a)$ and $\ell=b \tan v+\mathcal{O}(r_k,a)$. Indeed, the integral of geodesic curvature could be calculated in the following way
\begin{align}
\int_S^R\kappa_gd\ell=&-\frac{a\,{r_k} }{b^2}\int_{\phi_S}^{\phi_R}\cos v\,dv -\frac{a\,\alpha\,{r_k}}{b^2}\int_{\phi_S}^{\phi_R}\Big[\cos v +\ln ({r_k}) \cos v] dv, 
\notag\\
=&\frac{a\,{r_k} }{b^2}\Big[\sqrt{1-b^2{u_R}^2}+\sqrt{1-b^2{u_S}^2}\Big] 
\notag\\
+&\frac{a\,{r_k}(\alpha-1) }{b^2}\left(1-\ln(r_k)\right)\Big[\sqrt{1-b^2{u_R}^2}+\sqrt{1-b^2{u_S}^2}\Big]+\mathcal{O}(r_k^2,\alpha^2).
\label{int-kappag-second}
\end{align}
Considering  the  infinite distance limit $u_S,u_R\to 0$,  the total  expression of the deflection angle   of light is expressed as follow 
\begin{equation}
\Theta = \frac{2rk}{b}-\frac{2\,a\,r_k}{b^2}-\frac{r_k\,(\alpha-1)}{b}\left(1-2\ln(r_k)\right)-\frac{2 a\,{r_k}(\alpha-1) }{b^2}\left(1-\ln(r_k)\right). 
\end{equation}
Concretely, for $r_k=2M$ and $\alpha=1$ we  recover   the deflection angle of Kerr solution. However, for $\alpha\neq 1$  the deflection angle   is controlled by  the relevant parameter associated with the RKK solution including  Kiselev radius.  Indeed, we illustrate the deflection angle behaviors in Fig(\ref{aa}) by varying the rotation,  Kiselev radius and $\alpha$ parameters. 
\\
A close examination show that the plotted quantity  decrease by increasing the rotation $a$ or  $\alpha$.  However, this angle increase with the  Kiselev radius $r_k$.  For larger  values of $r_k$ the angle of deflection of RKK solution is larger  than Kerr black hole. However,  for small ones, the light deflection by Kerr solution is bigger than RRK black hole.  It is noted that, the deflection angle behavior of RRK is different than Kerr one by varying $a$ or $\alpha$. In compassion with serval works, this angle is similar to the angle of deflection of the quintessential rotating black holes\cite{Sharzz}. 
\section{Conclusion}

In this paper, we have derived a rotating solution of the reduced-Kiselev black hole through the \text{modified} Newman-Janis procedure. Then, we have examined  the horizon  and ergosphere geometries for such a solution. By inspecting the inner and outer horizon regions as a function of the involved parameters, it has been showed that the proportion of the two regions increases with $r_k$. Moreover, we have found   that   the gap between the inner and outer horizon increases significantly with $r_k$. It has also been remarked that the outer ergosphere becomes more stretched by increasing the values of $r_k$ or $a$ parameters. Concerning the shadow, we have observed the RKK black hole could keep a similar shadow to Kerr solution for particular values of the spin parameter $a$. Regarding the parameter $r_k$, we have addressed that it governs the shadow size while the parameters $\alpha, a$ are responsible of the shadow shape. With the use of the EHT shadow image of $M87^*$, we have constrained the involved parameter and have obtained a perfectly similar shadow image. Then, we have determined the emission rate expression and discussed the different feature of such a quantity. Particularly, we have showed that the latter increases proportionally to the parameter $\alpha$ while the opposite effect is observed for the spin $a$. Regarding the parameter $r_k$, we have noticed that such a quantity increases with the latter until it reaches a certain maximum and start decreasing again. Finally, we have computed the angle of deflection in the vicinity of RRK black holes. We have found   that  the deflection of light decreases by increasing either the spin $a$ or the parameter $\alpha$. However, such angle has increased with the $r_k$ radius values. Moreover,  we also have noticed that this angle is similar to the angle of deflection of a rotating black holes with quintessence\cite{Sharzz}.

In summary, the present solution has provided various distinctions and similarities compared to the other cases. For instance, we have obtained an elliptical shadow geometry in contrast to ordinary black hole solutions for small values of $\alpha$. Besides, we have observed that  the deflection angle behavior of RRK is different than Kerr one by varying $a$ or $\alpha$. Conversely, we have found that the shadow   can be compatible to the Kerr black hole and to $M87^*$ shadows. Moreover, an essential feature has been discovered when analysing the shadow shapes. Indeed, by keeping the ratio $a/r_k$ constant, we remark that the shadow shapes are the same while the size may be different depending on the value of $r_k$. Such similarities and distinctions push one to wonder whether the shadow images and deflection of light can be used to confirm or cancel the different models in literature. We hope to find relevant approaches to adress such issue in our future work.
\section*{Acknowledgements}
The authors would like to thank Benyounes Bel Moussa,  Hasan El Moumni,  El Houssaine El Rhaleb, Yassine Hassouni, Mustapha Lamaaoune, Alessio Marrani, Mohamed Oualaid, Moulay Brahim Sedra and Emilio Torrente-Lujan for discussions and collaborations on related topics\\
 The authors would like to express their heartfelt gratitude to  Kaoutar BENALI, Khaoula,  Jim EL BALALI  for their unwavering support and invaluable advice. 
\begin{appendix}
\end{appendix}

\end{document}